\documentclass[
 reprint,
superscriptaddress,
amsmath,amssymb,
aps,
]{revtex4-1}

\usepackage{graphicx}
\usepackage{amssymb}
\usepackage{dcolumn}
\usepackage{bm}
\usepackage{xcolor}
\usepackage{todonotes}
\usepackage{hyperref}
\hypersetup{
    colorlinks=true,
    linkcolor=blue,
    filecolor=magenta,      
    urlcolor=red,
    citecolor=blue,
}

\begin{document}

\preprint{phonons in O K edge RIXS}

\title{Doping-dependence of the electron-phonon coupling in two families of bilayer superconducting cuprates}

\author{Yingying Peng}
\email{yingying.peng@pku.edu.cn}
\affiliation{International Center for Quantum Materials, School of Physics, Peking University, Beijing CN-100871, China}

\author{Leonardo Martinelli}
\email{leonardo.martinelli@polimi.it}
\affiliation{Dipartimento di Fisica, Politecnico di Milano, piazza Leonardo da Vinci 32, I-20133 Milano, Italy}

\author{Qizhi Li}
\affiliation{International Center for Quantum Materials, School of Physics, Peking University, Beijing CN-100871, China}

\author{Matteo Rossi}
\affiliation{Dipartimento di Fisica, Politecnico di Milano, piazza Leonardo da Vinci 32, I-20133 Milano, Italy}

\author{Matteo Mitrano}
\altaffiliation[Present address:  ]{Department of Physics, Harvard University, Cambridge, Massachusetts 02138, USA}
\affiliation{Department of Physics and Materials Research Laboratory, University of Illinois, Urbana, IL 61801, USA\looseness=-1} 

\author{Riccardo Arpaia}
\affiliation{Dipartimento di Fisica, Politecnico di Milano, piazza Leonardo da Vinci 32, I-20133 Milano, Italy}
\affiliation{Quantum Device Physics Laboratory, Department of Microtechnology and Nanoscience, Chalmers University of Technology, SE-41296 G\"oteborg, Sweden\looseness=-1}

\author{Marco Moretti Sala}
\affiliation{Dipartimento di Fisica, Politecnico di Milano, piazza Leonardo da Vinci 32, I-20133 Milano, Italy}

\author{Qiang Gao}
\affiliation{Beijing National Laboratory for Condensed Matter Physics, Institute of Physics, Chinese Academy of Sciences, Beijing CN-100190, China}

\author{Xuefei Guo}
\affiliation{Department of Physics and Materials Research Laboratory, University of Illinois, Urbana, IL 61801, USA} 

\author{Gabriella Maria De Luca}
\affiliation{Dipartimento di Fisica Ettore Pancini Università di Napoli "Federico II",
Complesso Monte-Santangelo via Cinthia, I-80126 Napoli, Italy}
\affiliation{CNR-SPIN Complesso Monte-Santangelo via Cinthia, I-80126 Napoli, Italy}

\author{Andrew Walters}
\affiliation{Diamond Light Source, Harwell Campus, Didcot OX11 0DE, United Kingdom}

\author{Abhishek Nag}
\affiliation{Diamond Light Source, Harwell Campus, Didcot OX11 0DE, United Kingdom}

\author{Andi Barbour}
\affiliation{National Synchrotron Light Source II, Brookhaven National Laboratory, Upton, NY 11973, USA}

\author{Genda Gu}
\affiliation{Condensed Matter Physics \rm{\&} Materials Science Department, Brookhaven  National Laboratory, Utpton, NY 11973, USA\looseness=-1}

\author{Jonathan Pelliciari}
\affiliation{National Synchrotron Light Source II, Brookhaven National Laboratory, Upton, NY 11973, USA}

\author{Nicholas B.~Brookes}
\affiliation{ESRF—The European Synchrotron, 71 Avenue des Martyrs, CS 40220, F-38043 Grenoble, France}

\author{Peter Abbamonte}
\affiliation{Department of Physics and Materials Research Laboratory, University of Illinois, Urbana, IL 61801, USA} 

\author{Marco Salluzzo}
\affiliation{CNR-SPIN Complesso Monte-Santangelo via Cinthia, I-80126 Napoli, Italy}

\author{Xingjiang Zhou}
\affiliation{Beijing National Laboratory for Condensed Matter Physics, Institute of Physics, Chinese Academy of Sciences, Beijing CN-100190, China}

\author{Ke-Jin Zhou}
\affiliation{Diamond Light Source, Harwell Campus, Didcot OX11 0DE, United Kingdom}

\author{Valentina Bisogni}
\affiliation{National Synchrotron Light Source II, Brookhaven National Laboratory, Upton, NY 11973, USA\looseness=-1}

\author{Lucio Braicovich}
\affiliation{Dipartimento di Fisica, Politecnico di Milano, piazza Leonardo da Vinci 32, I-20133 Milano, Italy}
\affiliation{ESRF—The European Synchrotron, 71 Avenue des Martyrs, CS 40220, F-38043 Grenoble, France\looseness=-1}

\author{Steven Johnston}
\email{sjohn145@utk.edu}
\affiliation{Department of Physics and Astronomy, The University of Tennessee, Knoxville, TN 37996, USA}
\affiliation{Institute for Advanced Materials and Manufacturing, University of Tennessee, Knoxville, TN 37996, USA\looseness=-1}

\author{Giacomo Ghiringhelli}
\email{giacomo.ghiringhelli@polimi.it}
\affiliation{Dipartimento di Fisica, Politecnico di Milano, piazza Leonardo da Vinci 32, I-20133 Milano, Italy}
\affiliation{CNR-SPIN, Dipartimento di Fisica, Politecnico di Milano, piazza Leonardo da Vinci 32, I-20133 Milano, Italy\looseness=-1}

\date{\today}

\begin{abstract}
While electron-phonon coupling (EPC) is crucial for Cooper pairing in conventional superconductors, its role in high-$T_c$ superconducting cuprates is debated. Using resonant inelastic x-ray scattering at the oxygen $K$-edge, we studied the EPC in Bi$_2$Sr$_2$CaCu$_2$O$_{8+\delta}$ (Bi2212) and Nd$_{1+x}$Ba$_{2-x}$Cu$_3$O$_{7-\delta}$ (NBCO) at different doping levels ranging from heavily underdoped ($p =0.07$) to overdoped ($p=0.21$). We analyze the data with a localized Lang-Firsov model that allows for the coherent excitations of two phonon modes.
While electronic band dispersion effects are non-negligible, we are able to perform a study of the relative values of EPC matrix elements in these cuprate families.
In the case of NBCO, the choice of the excitation energy allows us to disentangle modes related to the CuO$_3$ chains and the CuO$_2$ planes. Combining the results from the two families, we find the EPC strength decreases with doping at $\mathbf{q_\parallel}=(-0.25, 0)$ r.l.u., but has a non-monotonic trend as a function of doping at smaller momenta. This behavior is attributed to the screening effect of charge carriers. We also find that the phonon intensity is enhanced in the vicinity of the charge-density-wave (CDW) excitations while the extracted EPC strength appears to be less sensitive to their proximity. By performing a comparative study of two cuprate families, we are able to identify general trends in the EPC for the cuprates and provide experimental input to theories invoking a synergistic role for this interaction in $d$-wave pairing.
\end{abstract}

\maketitle

\section{Introduction} 
The pairing interaction in conventional superconductors is mediated by phonons and critically depends on the electron-phonon coupling (EPC) \cite{BCS, McMillan}. The role played by EPC in copper-based high critical temperature superconductors (HTSs), however, is still debated. On the one hand, experimental probes like angle-resolved photoemission spectroscopy (ARPES) \cite{Lanzara2001, PhysRevLett.93.117003, DevereauxPRL2004, Cuk2005, Graf2008, HeScience, Johnston2012} and scanning tunneling microscopy (STM) \cite{Lee2006, PhysRevLett.103.236403, PhysRevB.75.214507} have found evidence for the interplay between lattice vibrations and superconductivity. On the other hand, spin fluctuations are widely believed to provide the dominant contribution to pairing in the cuprates \cite{Bulut1996, MaierPRL2008, Scalapino2012}, since the superconducting order parameter is of $d_{x^2-y^2}$ symmetry ~\cite{RevModPhys.72.969} and superconductivity appears in proximity to antiferromagnetism. This observation, however, does not explicitly rule out a potential role for phonons in pairing. For example, while a complete microscopic picture is currently absent, the interplay of the electron-phonon and electron-electron interactions may lead to the enhanced critical temperature ($T_c$) in HTSs. Indeed, theoretical work has shown that the materials with the largest coupling to the out-of-phase bond-buckling oxygen phonon branch (often referred to as the ``$B_{1g}$" modes) also have the largest $T_{c,\mathrm{max}}$ \cite{Johnston2010}, suggesting that these modes are relevant to superconductivity. 

HTSs have complex phase diagrams that include competing spin-, charge- and pair-density-waves, a pseudogap, and strange metallic behavior \cite{RevModPhys.87.457, Keimer2015, Comin2016, Pdw_review, Arpaia2018, arpaia2021charge, wahlberg2021restored}. 
The pivotal parameter that determines the nature of the ground state at a given temperature is the doping $p$, which gives the number of holes introduced in the CuO$_2$ planes. In the underdoped region of the phase diagram, the phonon branches are also renormalized in the presence of charge-density-wave (CDW) correlations. This behavior is reminiscent of what happens in conventional metals, where the EPC is the driving force behind the formation of charge modulations \cite{Zhu2367}. Recent resonant inelastic X-ray scattering (RIXS) studies have indeed observed anomalous softening and enhanced phonon excitations in the vicinity of charge order \cite{Chaix2017, Miao2018, lin2020nature, LiPNAS2020, PengLESCO}.

In the underdoped region of the phase diagram the coupling between phonon and electrons is also affected by the onset of the pseudogap phase. The pseudogap in HTSs is characterized by a temperature $T^*$, below which a partial gap opens in the normal state \cite{timusk1999pseudogap}. In general, $T^*$ falls to zero at a critical doping $p_c=0.19$, where Hall coefficient measurements show a change of carrier density from $p$ to $1+p$ \cite{badoux2016change}. Interestingly, a recent ARPES study of   Bi$_2$Sr$_2$CaCu$_2$O$_{8+\delta}$ (Bi2212) revealed a rapid change of bosonic coupling strength to the bond-buckling $B_{1g}$ phonon branch ($\sim 37$ meV) across the pseudogap boundary \cite{HeScience}. To better understand these observations, it is necessary to investigate how EPC changes with doping across the cuprate phase diagram, spanning from the underdoped region to the overdoped Fermi liquid region found for $p > p_\text{c}$.

Measuring the full EPC vertex $M({\bf k},{\bf q})$ is challenging, as it can depend on both the fermionic (${\bf k}$) and bosonic (${\bf q}$) momenta. Techniques like ARPES probe the ${\bf k}$ dependence of the EPC via the electron self-energy, which averages over the phonon momentum ${\bf q}$. Similarly, scattering probes like inelastic x-ray and neutron scattering tend to probe the ${\bf q}$ dependence of $M({\bf k},{\bf q})$ via the phonon line width, which averages over the fermion momentum ${\bf k}$. Recently, RIXS has been demonstrated as an effective tool to determine the momentum dependence of the EPC with element specificity, due to the interaction between the excited electrons in the intermediate state and phonons \cite{Ament2011}. As a scattering technique, RIXS more directly measures the ${\bf q}$-dependence of the vertex and integrates over the ${\bf k}$-dependence. For example, models have shown that the ${\bf k}$-integration can be weighting by other factors like the electron occupations and orbital character of the bands if the system hosts itinerant electrons~\cite{Ament2011, Devereaux2016, braicovich2020determining}. In cases where the electron mobility can be neglected (e.g. due to a strong core-hole potential), the EPC strength can be extracted either from the ratio between the intensity of the phonon and of its overtones in RIXS spectra \cite{Ament2011, bieniasz2020beyond}. Alternatively, one can measure the phonon intensity as the incident beam detuned from the resonance, as successfully done recently at the Cu $L_3$ in cuprates~\cite{braicovich2020determining, Rossi2019}.

Prior RIXS studies of EPC in the 2D cuprates have focused largely on the Cu $L_3$-edge \cite{Rossi2019, braicovich2020determining, PengLESCO}. However, it has been established that the bands that cross the Fermi level in doped cuprates have significant oxygen $2p_{x,y}$ orbital character \cite{chen1991electronic}. Moreover, the phonons believed to be most relevant to pairing in the cuprates are the optical oxygen modes~\cite{Cuk2005,Johnston2010}. It is, therefore, desirable to also study the EPC at the O $K$-edge in these materials. The RIXS process at this edge involves the $1s \rightarrow 2p$ transition, so that in the intermediate state the charge perturbation mostly involves the electron in a $2p$ orbital (plus a core hole, on whose influence we will discuss in the following). Moreover, RIXS experiments at the O $K$-edge can benefit from a better energy resolution and longer-lived core-hole compared to experiments conducted at the Cu $L_3$-edge \cite{Brookes2018}. 

Here, we have studied the coupling between electrons and the oxygen-related phonons in bi-layer compounds Bi2212 and NBCO, using O $K$-edge RIXS with an energy resolution of about 33 meV. We have extracted the EPC strength from the data using a localized model which is based on a Lang-Firsov transformation (we refer to the model as Lang-Firsov model in the following) \cite{LangFirsov, Geondzhian2020, STOPolarons} that accounts for both bond-buckling and bond-stretching phonons. Combining the results of the two cuprate families, we find that EPC strength decreases with doping at $\mathbf{q_\parallel}=(-0.25, 0)$ r.l.u. while this trend becomes non-monotonic at smaller momenta. The observed changes in the EPC with doping agree with the nonuniform screening effect discussed in theoretical calculations \cite{Johnston2010}. We have also studied the effect of CDW correlations on EPC. Our results provide experimental insights on the general trends of EPC and allow us to better understand their role in high-$T_c$ cuprate superconductors.

\section{Methods}
\subsection{RIXS experiments}
We have performed O $K$-edge RIXS studies on NBCO and Bi2212 samples with different doping levels. 
The NBCO samples were 100 nm films grown on SrTiO$_3$, with two different doping levels: a heavily underdoped sample with $T_\mathrm{c} = 38$~K (UD38, $p \simeq 0.07$) and an optimally-doped sample with $T_c = 92$ K (OP92, $p \simeq 0.16$). Lattice constants were $a=b=3.9~\textrm{\AA}$, while $c = 11.714~\textrm{\AA}$ in the UD38 sample and $c=11.74~\textrm{\AA}$ in the OP one. RIXS spectra were collected at the I21 beam line of the Diamond Light Source. 
Energy resolution was determined to be $34$ meV by measuring the width of the elastic line on a carbon tape. We used a fixed temperature of $20$~K.

The Bi2212 samples were single crystals with three doping levels: optimally doped with $T_c = 91$~K (OP91, $p\simeq 0.16$), overdoped with $T_c = 82$~K (OD82, $p\simeq 0.19$), and overdoped with $T_c = 66$~K (OD66, $p\simeq 0.21$). The RIXS measurements on Bi2212 were performed using the soft x-ray inelastic beamline (SIX) $2$-ID at the National Synchrotron Light Source II, Brookhaven National Laboratory \cite{bnlbeamline}. 
The total experimental energy resolution was $\sim 33$ meV. We measured at $35$~K to avoid sample damage. We define our reciprocal lattice units (r.l.u.) in terms of the tetragonal unit cell with $a=b=3.8~\textrm{\AA}$ and $c=30.89~\textrm{\AA}$. 

The RIXS spectra were collected with $\sigma$ incident polarization (perpendicular to the scattering plane) to maximize the charge signal. The scattering angle was fixed at 150$^\circ$, and we used a grazing-in geometry, which by convention means negative parallel transferred momentum $\mathbf{q_\parallel}$, as shown in Fig.~\ref{fig:rawdata}(g).

\subsection{The Lang-Firsov Localized Models}

\label{sec:amentGilmore}
The first model developed to describe the EPC in RIXS experiments was the single-site model introduced in Ref.~\cite{Ament2011} to describe lattice excitations generated at transition metal $L$-edges. The model was derived starting from the local Hamiltonian
\begin{equation}
{\cal H}=\epsilon_0 n_d + \omega_0b^\dagger b + M \sum_\sigma n_d (b+b^\dagger),
\end{equation}
where $d^\dagger_\sigma$ and $b^\dagger$ represent the creation operators of a spin $\sigma$ electron and local phonon mode, respectively, $n_d = \sum_\sigma d^\dagger_\sigma d^{\phantom\dagger}_\sigma$, 
and $M$ is the EPC constant, which is related to a dimensionless coupling parameter $g = (M/\omega_0)^2$. This model is based on the simplifying assumption that both the electrons and phonon are \emph{local} in the sense that they cannot travel through the lattice at any stage of the RIXS process. 
In this limit, the RIXS intensity can be computed exactly by use of a Lang-Firsov transformation, yielding 
\begin{equation}\label{eqn:ament}
    I(q,\Omega) = \sum_{n}|A_n(\Delta+\mathrm{i}\Gamma)|^2 \delta(\omega_0n-\Omega), 
\end{equation}
where $\Delta$ is the detuning of the incident energy from the resonant peak, $\Gamma$ is the half-lifetime of the intermediate state of RIXS process, and 
\begin{equation}
    A_n(z) = \sum_m \frac{B_{n,m}(g)B_{m,0}(g)}{z-\omega_0(m-g)} 
\end{equation} 
with 
\begin{equation}
B_{n,m} = (-1)^n \sqrt{e^{-g}n!m!}\sum_{l=0}^m \frac{(-g)^l g^{\frac{n-m}{2}}}{(m-l)!l!(n-m+l)!}
\end{equation}
being the standard Franck-Condon factor ($B_{n,m}$ is understood to mean 
$B_{\mathrm{max}(n,m),\mathrm{min}(n,m)}$). 

Reference \cite{Geondzhian2020} recently generalized the single-site model to the case where the localized electron couples to more than one type of 
phonon. This generalization allows each phonon to be coherently excited and is relevant to the cuprates where we naturally expect coupling to different phonon branches. Here, we consider the case with two types of phonon modes (e.g. the bond-buckling and bond-stretching modes),  again described by the local Hamiltonian \cite{Geondzhian2020}
\begin{eqnarray}
{\cal H}=~&&\epsilon_0 n_d +
\sum_{\lambda=1,2}
\left[\omega^{\phantom\dagger}_\lambda b_\lambda^\dagger b^{\phantom\dagger}_\lambda+
M_\lambda n_d(b^{\phantom\dagger}_\lambda+b_\lambda^\dagger)\right].
\end{eqnarray}
The exact RIXS intensity can still be computed, yielding 
\begin{equation}\label{eqn:gilmore}
    I(q,\Omega) = C_\mathrm{scale} \sum_{n_1,n_2} |A_{n_1,n_2}(\Delta+\mathrm{i}\Gamma)|^2\delta
    \big( \sum_{\lambda=1,2} n_\lambda \omega_\lambda - \Omega\big), 
\end{equation}
where 
\begin{eqnarray}\label{eqn:gilmore2}
{A_{n_1n_2}} = \sum_{m_1m_2}
\frac{D_{n_1m_1}^{n_2m_2}(g_1,g_2)D_{m_10}^{m_20}(g_1,g_2)}
{\Delta + i\Gamma-\sum_{\lambda=1,2}\omega_\lambda(m_\lambda-g_\lambda)}.
\end{eqnarray}
Here, $C_\mathrm{scale}$ is an overall amplitude factor, independent of momentum and detuning and applies to all phonon branches, which we use to scale overall intensity to match experiment. The coefficients $D$ in Eq.~(\ref{eqn:gilmore2})  are given by a product of Franck-Condon factors
\begin{equation}
    D_{n_1,m_1}^{n_2,m_2}(g_1,g_2) = B_{n_1,m_1}(g_1)B_{n_2,m_2}(g_2), 
\end{equation}
where $g_\lambda = (M_\lambda/\omega_\lambda)^2$.  

We stress that both models neglect all electron mobility involved in the RIXS scattering process. 
Nevertheless, relatively simple expressions for the RIXS intensity [Eqs. (\ref{eqn:ament}) and (\ref{eqn:gilmore})] are obtained from these approximations that are useful for fitting experimental data. The parameters setting the intensity of phonon excitations in both models are the EPC strength of the different branches $g_\lambda$, the detuning energy $\Delta$, and the core-hole lifetime parameter $\Gamma$. Throughout, we treat $g_\lambda$ as a fitting parameter and set $\Delta$ based on experimental conditions. We will also fix $\Gamma = 0.15$ eV \cite{Lee2013, Johnston2016, braicovich2020determining} for all the samples investigated, since the intrinsic lifetime of the state is mostly determined by Auger recombination rate and so independent of the material studied. 

A recent theoretical study in the single carrier limit~\cite{bieniasz2020beyond} found that the local model for the RIXS cross-section produces stronger phonon features in comparison to a fully itinerant model. Moreover, the discrepancy between the models becomes more pronounced as the strength of the localizing core-hole potential is reduced. This result suggests that fitting a localized model to phonon spectra may \emph{underestimate} the strength of the EPC. In the single carrier limit, electron mobility in the intermediate state can also introduce a momentum dependence in the intensity of the phonon excitations that is unrelated to the momentum dependence of the EPC constant $M({\bf k},{\bf q})$. 
With these caveats in mind, we proceed by fitting the experimental data with the localized models with the assumption that doing so will allow us to extract general trends in the data. For these reasons, the absolute values of the extracted couplings obtained here should not be equated to the value entering a microscopic Hamiltonian. 

\section{Experimental results}

\begin{figure*}
\includegraphics[width = \textwidth]{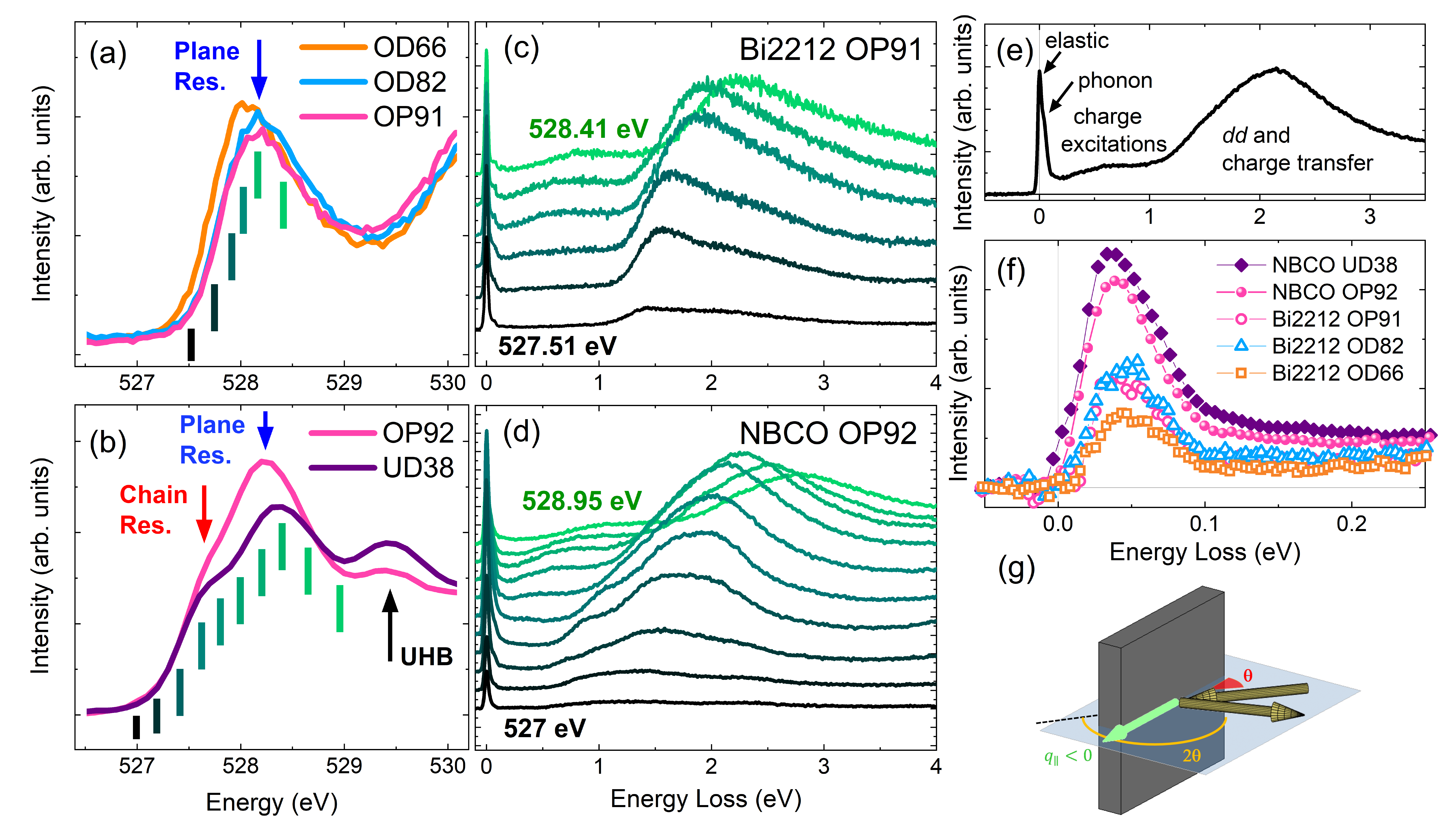}
\caption{\label{fig:rawdata} Overview of data on NBCO and Bi22212 at the O $K$-edge. (a, b) XAS spectra for different doping values of Bi2212 and NBCO, respectively. The legends indicate the critical temperature of the samples. The relevant resonances are indicated with blue and red arrows. (c, d) Overview of RIXS spectra in Bi2212 OP91 and NBCO OP92, respectively. The excitation energies for the different spectra are highlighted as ticks in the XAS, using the same colour code. (e) Example of RIXS spectrum at Oxygen $K$-edge collected on NBCO OP92 at 528.2 eV. Labels indicate the different excitations. (f) Zoom-in of the low-energy regions of RIXS spectra for all the samples acquired at $H=-0.15$ r.l.u. and at the energy corresponding to the plane resonance. The elastic peak has been subtracted for clarity. The spectra have been normalized to the spectral weight between 1 and 7 eV corresponding to the CT excitation range. (g) Sketch of the experimental geometry. The incoming and outgoing x-rays are reported in yellow arrows, and the incident ($\theta$) and scattering ($2\theta$) angles are highlighted. The component of the transferred momentum parallel to the CuO$_2$ planes is reported in light green.}
\end{figure*}

\subsection{\label{sec:rawdata}  RIXS spectra at O $K$-edge}

\begin{figure*}[htbp]
\includegraphics[width = \textwidth]{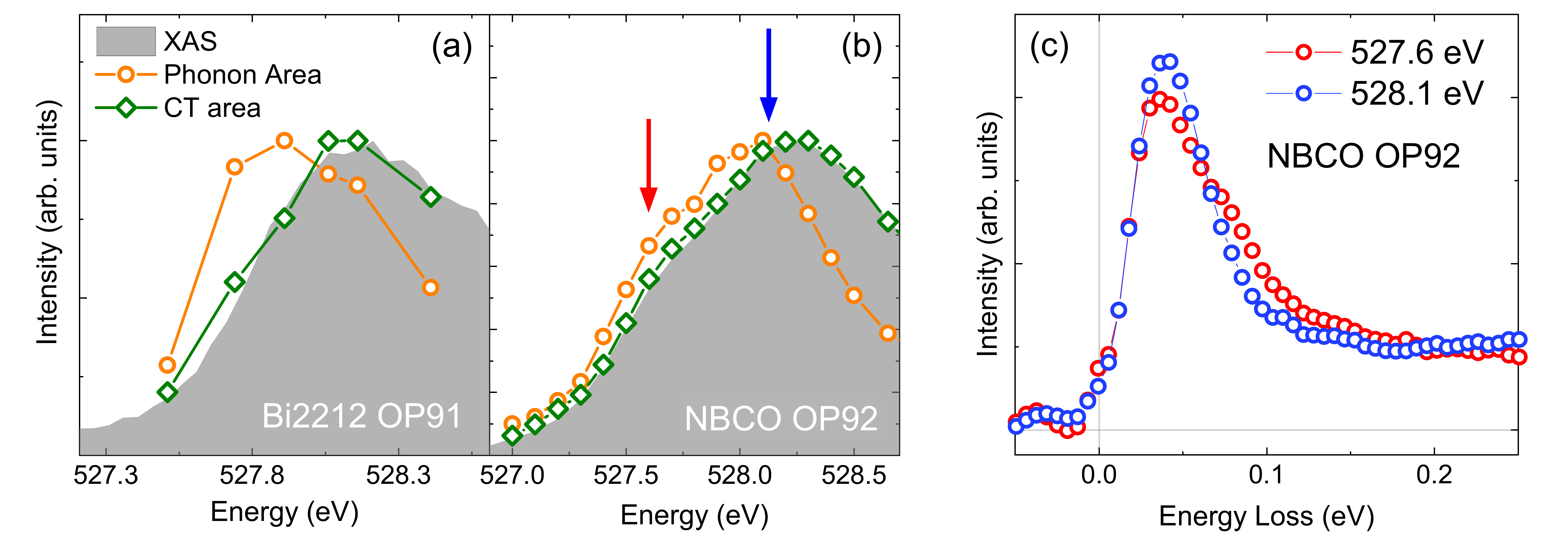}
\caption{\label{fig:resonancesNd123Bi2212} XAS and phonon resonances in Bi2212 OP91 and NBCO OP92. (a, b) XAS spectra (grey shade area), the spectral area of phonon (orange circles) and charge-transfer excitations (green diamonds) for Bi2212 OP91 and NBCO OP92, respectively. The red and blue dashed vertical lines indicate the chain and plane resonant energies in NBCO OP92, respectively. (c) RIXS spectra collected at the chain (red) and plane (blue) resonances in NBCO OP92 after subtracting the elastic peaks for comparison.
}
\end{figure*}

The X-ray absorption spectra (XAS) of our Bi2212 and NBCO samples are displayed in Fig.~\ref{fig:rawdata}(a) and Fig.~\ref{fig:rawdata}(b), respectively. It is easy to identify the Zhang-Rice singlet (ZRS) pre-peak in the O $K$-edge XAS of Bi2212, which originates from the hybridization of the in-plane O $2p_\sigma$ and Cu $3d_{x^2-y^2}$ orbitals \cite{hawthorn2011resonant}. Both the energy and amplitude of the ZRS peak change with doping as expected. On the other hand, we observe two different structures in NBCO whose intensity increases with doping: a main peak at 528.2 eV, and another peak at a lower energy ($\sim 527.6$ eV). The former one is associated with doped holes in the CuO$_2$ planes, while the latter originates from the CuO$_3$ chains, which act as a charge reservoir in NBCO~\cite{hawthorn2011resonant}. Our UD38 sample has short chains in both in-plane directions due to the tetragonal structure, while the OP92 sample has fully developed long chains, and due to the twin structure, they exist in both in-plane directions.
The comparison between UD38 and OP92 sample also nicely highlights the decrease in spectral weight of the upper Hubbard band (UHB) peak around 529.5 eV.

To investigate the electron-phonon interaction, we have acquired RIXS spectra at different incident X-ray energies across the doping-related resonances. Figs. \ref{fig:rawdata}\textcolor{blue}{(c)} and \ref{fig:rawdata}\textcolor{blue}{(d)} present an overview of the RIXS spectra for Bi2212 OP91 and NBCO OP92, respectively. The incident energy of each scan is indicated by the tick marks with the same color in panels (a) and (b) for the two compounds. As highlighted in Fig.~\ref{fig:rawdata}(e), we recognize four features in the spectra: the elastic peak, the phonons (and their overtones) below 0.2 eV, the broad charge excitations peaked around 0.8 eV, and some higher-energy excitations above 1.5 eV from $dd$ and charge-transfer (CT) excitations. Moreover, we also identify some spectral weight around 0.5 eV, which could be due to broad bimagnon excitations \cite{BisogniPRB2012}. 
To highlight the phonon contribution to the spectra and compare intensities between different samples at resonant energy, we removed the elastic peak and normalized the spectra to their respective \emph{dd}+CT  excitations as summarized in Fig.~\ref{fig:rawdata}(f). The phonon excitations have a larger spectral weight in NBCO and the exciations in Bi2212 OD82 ($p_c$=0.19) have a stronger intensity than those observed in the OP91 and OD66 samples.

To precisely determine the resonant energy of the phonon signal (i.e., the energy at which the phonon intensity is maximum), we have extracted their integrated spectral weight. This value was determined by removing the elastic peak and charge excitations from RIXS spectra, and then integrating the infrared region of the RIXS spectra ($0 - 300$~meV). As a reference, we have also extracted the spectral weight of CT excitations by integrating the spectra in a 3~eV window centered around their maximum value. The integrated spectral weight of the phonon and CT excitations are plotted in Figs.~\ref{fig:resonancesNd123Bi2212}\textcolor{blue}{(a)} and \ref{fig:resonancesNd123Bi2212}\textcolor{blue}{(b)} for Bi2212 OP91 and NBCO OP92, respectively, normalized to their maximum value and superimposed over the corresponding XAS spectra. We notice that the one-dimensional breathing mode of the chains has a stronger EPC strength than its two-dimensional counterpart, which is evident from the higher intensity of its first overtone. 

The intensity of CT excitations follows closely the XAS spectra in both samples, confirming the stability of our incident energy and the independence of the lifetime of the intermediate state. However, the phonon has a large shift of $200 - 250$~meV relative to the maximum of XAS, both in Bi2212 and NBCO. The presence and magnitude of this energy shift are puzzling. A small displacement, of the order of the phonon energy ($\approx 50 - 100$ meV) between the maximum of the XAS and the maximum of phonon intensity has been observed before \cite{feng2020Disparate, PengLESCO} and can be accounted for by the localized Lang-Firsov model with a strong EPC and a long core-hole lifetime. The large value observed here, however, requires a different explanation. As suggested in Ref.~\cite{Geondzhian2018}, a possible origin could be a change in the curvature of the potential energy surface between the ground and intermediate state of the RIXS process. 

\subsection{Energy detuning measurements} 

\begin{figure}[htbp]
\centering
\includegraphics[width=\columnwidth]{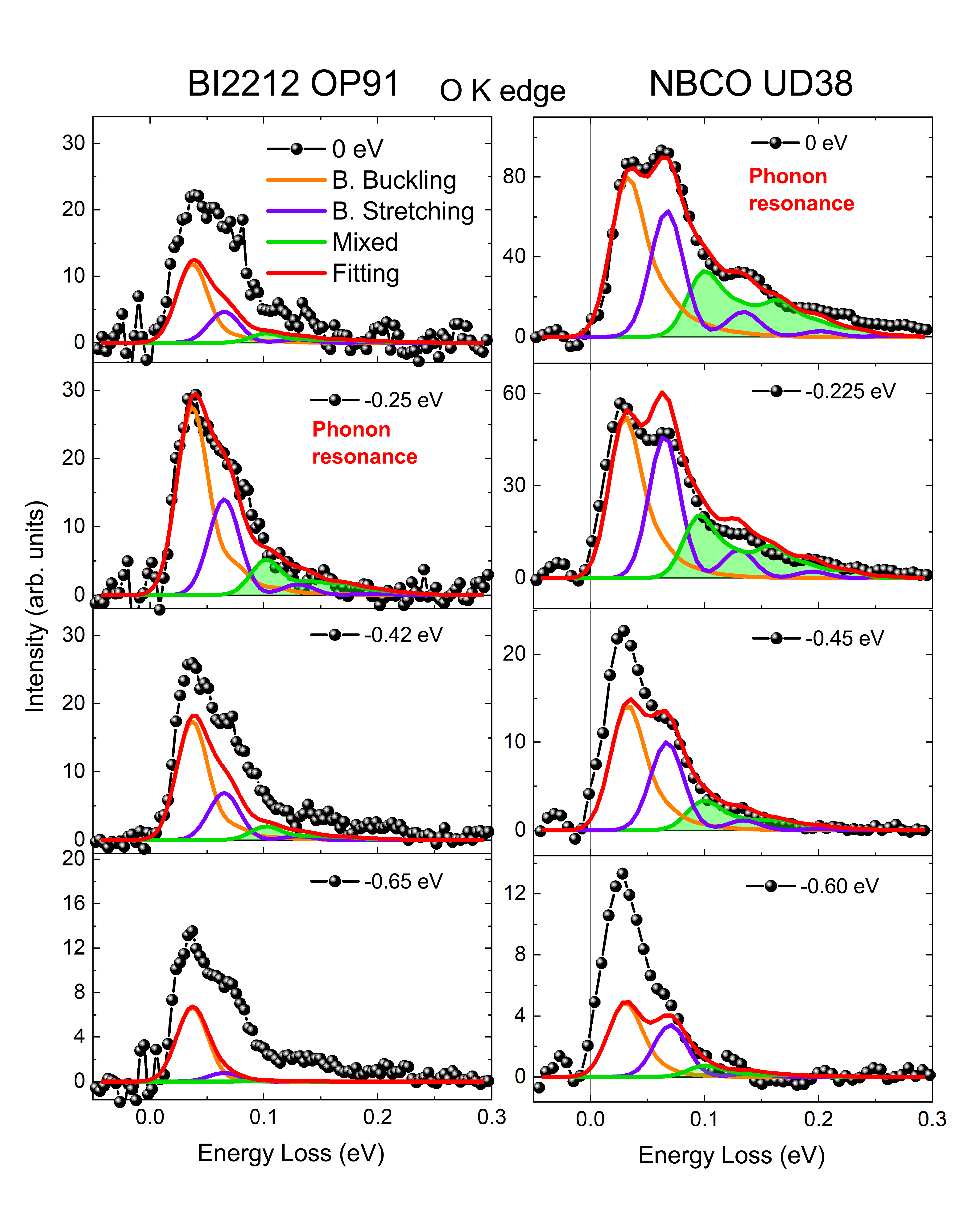} \\ 
\caption{\label{fig:detuning} Detuning data on Bi2212 OP91 (left panels) and NBCO UD38 (right panels). Incident energies were chosen around the plane resonance for Bi2212 OP91 and of chain resonance for NBCO AF, for the reasons discussed in the text. The momentum transfer was $(-0.15,0)$ r.l.u. for Bi2212 OP91 and $(-0.25,0)$ r.l.u. for NBCO AF. Solid lines are the fitting curves, decomposed into the various phonon branches, obtained through the global fitting discussed in the main text. We have highlighted that, in the case of plane resonance of Bi2212, the phonon resonance is shifted with respect to the maximum of the XAS. A failure of the model is evident at large detuning energies.} 
\end{figure}

We initially attempted to use the energy detuning method to extract the value of the EPC for the buckling and breathing branches, which has been described in detail in Ref.~\cite{braicovich2020determining} and has been successfully applied at the Cu $L_3$ edge~\cite{braicovich2020determining, Rossi2019}. This approach is based on the single-site model presented in Sec.~\ref{sec:amentGilmore} and consists in acquiring scans at different incident energies, at and below the resonant peak in the XAS. The EPC of the different phonon branches can then be extracted by fitting the intensities of the different phonon branches (and of their higher harmonics) with Eq.~\eqref{eqn:gilmore}. To reduce the number of parameters, we fit all of the spectra for a given sample at the same time, using a single overall scale factor and fixing $\Gamma = 0.15$~eV \cite{Lee2013, Johnston2016}. In this way, the only free parameters are the EPC strengths $M_\lambda$ of the two branches and their energies $\Omega_\lambda$. We have also subtracted the elastic peaks and considered only the inelastic parts, since the elastic peaks can carry additional contribution from other types of scattering.

As highlighted in Sec.~\ref{sec:rawdata}, one difficulty is that the maximum of the phonon signal is displaced from the maximum of XAS spectra by $\sim 200-250$~meV. While the employed model can account for a small displacement between the two energies (of the order of $\omega_\lambda$) in the strong coupling limit, the value of $M_\lambda$ that is needed is too large and would produce phonon harmonics inconsistent with the RIXS spectra. Therefore, we have decided to artificially put the zero of the detuning in the model at the phonon resonance. 

To test the validity of the detuning approach, we have used two different datasets: the chain resonance of NBCO OP92 and the plane resonance of Bi2212 OP91. We chose the chain resonance for NBCO for two reasons. First, it selects the one-dimensional CuO$_3$ chains as opposed to the two-dimesional CuO$_2$ planes of Bi2212, which allows us verify whether there are differences related to dimensionality. Secondly, the chain resonance energy ($527.6$ eV) is lower than that of the plane ($528.2$ eV). This difference means that the detuning energies of the chain are not obscured by additional contributions from the plane resonance, while the detuning energies of the plane overlap with the chains. 

The spectra with the corresponding fittings are presented in Fig. \ref{fig:detuning}. The solid orange and purple lines are the bond-buckling and bond-stretching phonons including the overtones, respectively, while the green shaded area represents the mixed coherent multi-phonons. However, it is evident from the spectra that the detuning method does not reproduce the measured intensities for all incident photon energies. While the agreement is quite good close to the resonance energy for both the cases, the calculated spectra far from the  resonance significantly underestimate (by more than $\sim 50$\%) the intensity of the phonon branches. This failure of the detuning method at the ligand $K$ edge is not completely unexpected given the very different nature of the intermediate state between the Cu $L_3$- and the O $K$-edges, and of the localized assumptions underpinning the Lang-Firsov models. Indeed, at the Cu $L_3$-edge, the excited electron is more strongly bound to the core hole in comparison to the O $K$-edge. This means that the excited electron is more free to propagate around the lattice during its lifetime in the latter case.  Reference~\cite{bieniasz2020beyond} recently developed a model for an intermediate state consisting of an itinerant electron interacting with phonons and the core hole in an otherwise empty band. One of the main results derived there is that the detuning curves are quantitatively different from the localized model; they have a weaker dependence on the EPC when the core-hole potentials are weak and they generally decay faster as a function of the detuning energy in comparison to the localized model. We observe a small difference between the decay of the intensity of the two phonon branches with detuning, as evident from Fig.~\ref{fig:detuning}. This observation suggests that the real situation close to the Fermi energy in cuprates is intermediate between the two approaches. 

\begin{figure*}[htbp]
\includegraphics[width = 0.9\textwidth]{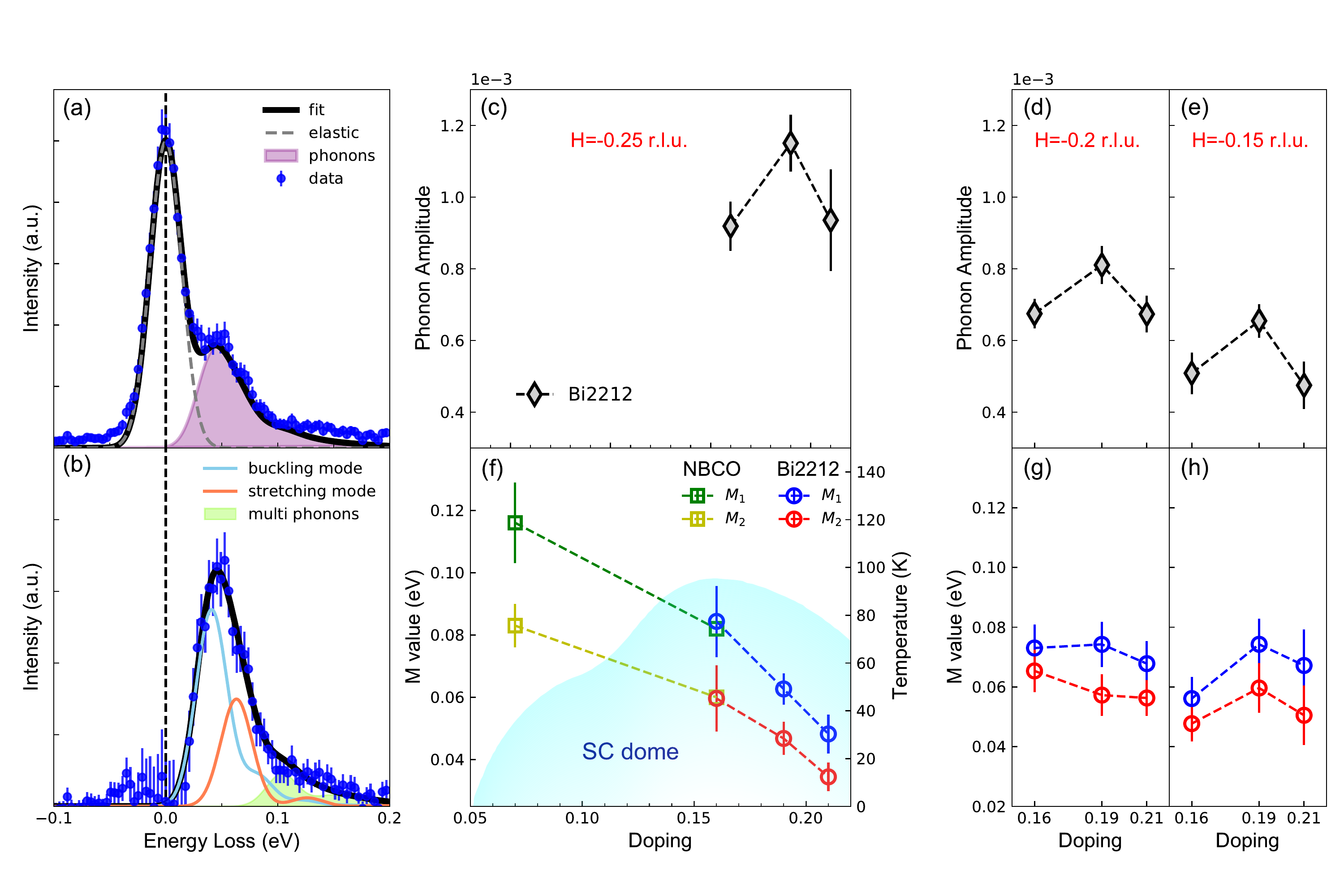}
\caption{\label{fig:dopingdep}(a) Fits of RIXS spectra after subtracting the high-energy excitations for Bi2212 OD82, taken at momenta transfer $H=-0.2$ r.l.u.. Elastic peak is fitted by a Gaussian function determined by the energy resolution. The phonon excitations are fitted with the Frank-Condon model considering two phonons, i.e. the buckling mode ($\sim 40$ meV) and stretching mode ($\sim 70$ meV). (b) Phonon components are shown after subtracting the elastic peak. (c-e) Doping dependence of phonon amplitude for Bi2212, which shows a maximum at $p_c=0.19$. (f-h) Doping dependence of the electron-phonon coupling strengths $M$ for both NBCO and Bi2212. The SC phase diagram \cite{SCdome} of cuprates (shaded area) is displayed in (f).}
\end{figure*}

\begin{figure}[htbp]
\flushleft
\includegraphics[width=0.6\textwidth]{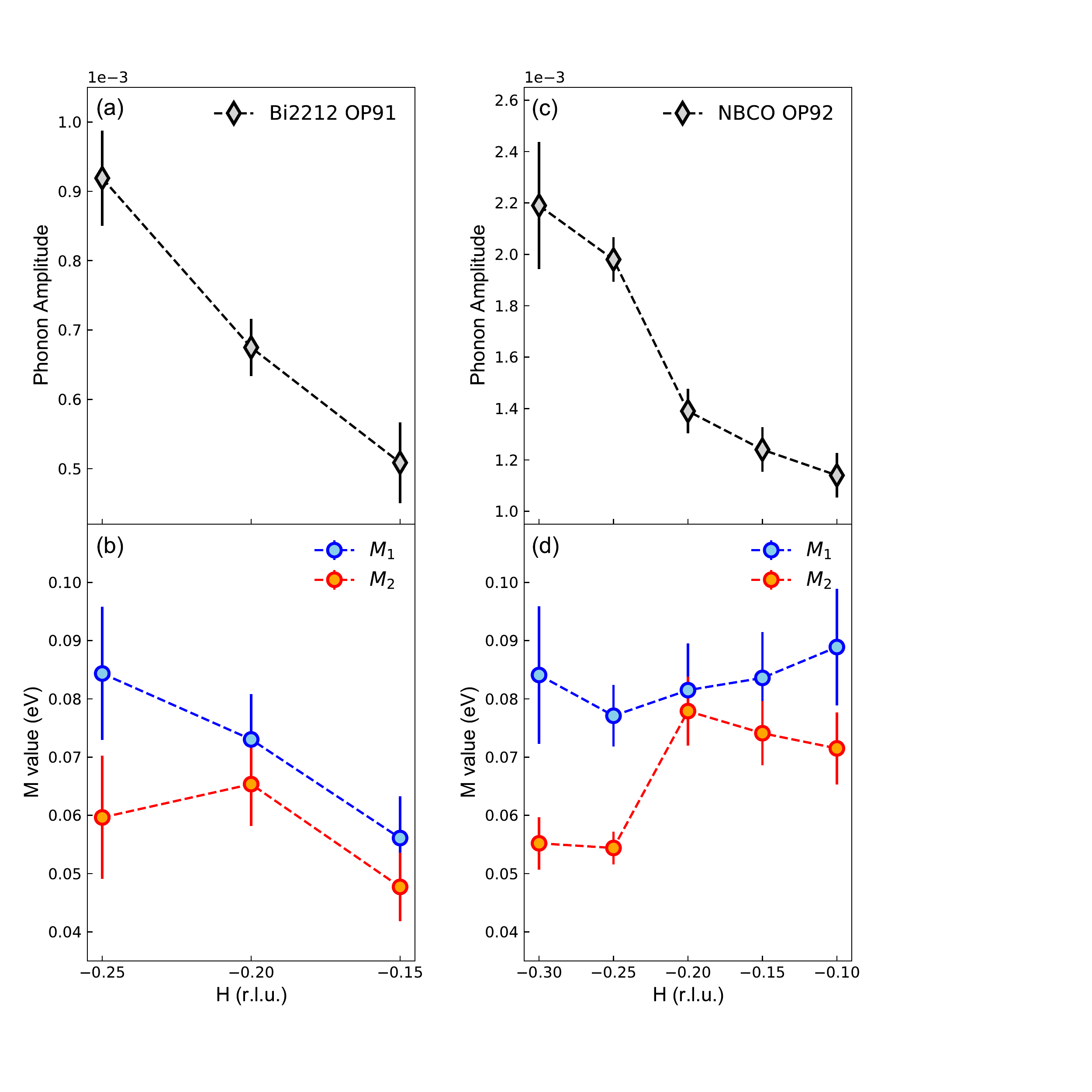}
\caption{\label{fig:momentumdep} The momentum dependence of the fitting parameters in Bi2212 OP91 (a,b) and NBCO OP92 (c,d). Phonon amplitudes increase with larger {\bf q}, while the $M$ values do not show a clear momentum dependence. The extracted $M$ values of two samples are close to one another. }
\end{figure}

\subsection{Effect of doping on EPC} 
Determining the EPC as a function of doping is crucial for understanding its potential contribution to $d$-wave pairing. For this purpose, we fit the RIXS spectrum at the resonant energy using Eq.~(\ref{eqn:gilmore}), after subtracting the elastic, charge, and CT excitations from the data, as shown in Fig.~\ref{fig:dopingdep}. As discussed above, the use of Eq.~(\ref{eqn:gilmore}) is likely to underestimate the overall strength of the EPC, particularly at the O $K$-edge, where electron itinerancy in the RIXS intermediate state can be more pronounced than at the Cu $L$-edge~\cite{bieniasz2020beyond}. We nevertheless proceed with Eq.~(\ref{eqn:gilmore}) in a comparative analysis to identify trends in the data, both as a function of doping and between Bi2212 and NBCO. Presumably, the degree of itinerancy in these systems is comparable for similar doping levels. Thus, the analysis for both samples should  be subject to the same systematic error when applying the local model to the data. One should keep in mind that the absolute magnitudes of EPC constants extracted in this analysis cannot be mapped directly onto the $M_\lambda({\bf q})$ entering in any microscopic Hamiltonian. 

To fit the data, we consider the coherent contribution from two phonon modes -- one from the bond-buckling branch ($\omega_1 \simeq 40$ meV) and one from the bond-stretching branch ($\omega_2 \simeq 70$ meV) -- while treating the coupling strengths $M_\lambda$ as fitting parameters. We also introduce a prefactor $C_\mathrm{scale}$, which multiplies the entire spectrum by a constant amount. We are able to obtain a consistent fit to the doping-dependent RIXS spectra using this approach. The doping evolution of the extracted amplitude $C_\mathrm{scale}$ for the Bi2212 sample are plotted in Figs.~\ref{fig:dopingdep}(c-e) for different momentum transfers, while the corresponding $M_\lambda$ values are displayed in Figs.~\ref{fig:dopingdep}(f-h). For comparison, we have also overlaid the doping dependence of the extracted $M_\lambda$ in NBCO at $H=-0.25$ r.l.u. in Fig.~\ref{fig:dopingdep}(f). Notably, the values of the extracted coupling in Bi2212 and NBCO are almost identical at optimal doping, which provides additional post-hoc justification for our comparative approach.  

For momentum transfer $H = -0.25$ r.l.u., we find that the EPC of the lower-energy buckling mode is stronger than the EPC of the breathing mode in both samples and for all the measured doping levels. This observation is consistent with the previous Cu $L$-edge RIXS measurements on NBCO UD38 sample for small momentum transfers \cite{braicovich2020determining} and prior ARPES experiments that infer the strongest coupling to the bond-buckling phonon modes \cite{PhysRevLett.93.117003, Cuk2005, Johnston2010, Johnston2012}. As the hole concentration increases, we find that the extracted $M$ values for both branches decrease monotonically at $H=-0.25$ r.l.u.. This behavior can be understood as being a consequence of the improved electronic screening as the system becomes more metallic \cite{Johnston2010}. The situation is quite different for smaller momenta. For $H=-0.20$ r.l.u., the inferred EPC constants are very weakly dependent on doping, while at 
$H = -0.15$ r.l.u. the doping dependence even becomes non-monotonic, with both branches becoming more strongly coupled near $p_c = 0.19$. 

The doping dependence summarized here in both compounds is consistent with the poor screening scenario discussed in Ref.~\cite{Johnston2010}. There, it was shown that the poor conductivity perpendicular to the CuO$_2$ plane results in a nonuniform screening of the electron-phonon interaction whereby it was well screened for large momentum transfers but poorly screened at small ${\bf q}$. The suppression of large ${\bf q}$ coupling relative to small ${\bf q}$ will increase the EPC in the $d$-wave channel, suggesting that the contribution of these modes to pairing increases as the doping increases and may become largest at optimal doping. 

\subsection{Influence of CDW on EPC}
Both optimally doped NBCO \cite{Ghiringhelli2012, BlancoCanosa2014} and Bi2212 \cite{hashimoto2014direct, daSilvaNeto2014, lee2021spectroscopic} present a CDW ordering with $q_\text{CDW} \simeq 0.29$ r.l.u., though the CDW signals become weaker with respect to underdoped samples \cite{BlancoCanosa2014}. Previous RIXS studies at Cu $L_3$-edge and O $K$-edge on doped cuprates have found an anomalous enhancement of phonon intensity near the CDW wavevector, which is ascribed to the presence of dispersive CDW excitations that overlap and interfere with the phonon excitations~\cite{Chaix2017,LiPNAS2020}. We have, therefore, searched for possible enhancements of EPC of buckling and breathing branches due to the presence of charge order. 

Employing again the generalized Lang-Firsov model [Eq.~(\ref{eqn:gilmore2})] that accounts for coupling to both modes, we have extracted the evolution of the EPC values and the overall scaling prefactor along the $(H,0)$ direction. Figure~\ref{fig:momentumdep} shows the momentum dependence of the phonon amplitudes ($C_\mathrm{scale}$) and $M_\lambda$ values in optimally doped Bi2212 and NBCO. Evidently, the phonon amplitudes increase monotonically on approaching the $q_\text{CDW}$. At the same time, the EPC does not show a clear trend with $M_1$ larger than $M_2$ for all our measured momenta. 
These results indicates that while the phonon excitations increase in intensity in proximity of the CDW, there is no clear corresponding enhancement in the relative intensity of the phonon overtones. This behavior can be reconciled if there is an overlapping contribution to the elastic line intensity due to the CDW excitations and no corresponding increase in the EPC close to ${\bf q}_\mathrm{CDW}$. One should, therefore, be careful in interpreting directly the phonon intensity as the EPC strength at O $K$-edge due to the longer intermediate state. The explanation of this discrepancy remains a topic for further investigation.\\

\section{Conclusions} 
We have measured the phonon dynamics in two families of cuprates superconductors as a function of incident energy, momentum and doping, ranging from strongly underdoped to overdoped samples. Our experimental results highlight that the localized Lang-Firsov model does not describe phonon spectra when the incident energy is detuned away from the resonance peak. We believe that this discrepancy is a fingerprint of an itineracy of the excited electron and phonon cloud, something which is not captured in the localized model \cite{bieniasz2020beyond}. The detuning method, which has been successfully applied at the Cu $L$-edge to extract the EPC, is therefore not applicable at the O $K$-edge within the present theoretical framework. 

By extracting the momentum and doping dependence of EPC measured at resonant x-ray energy, we have revealed a systematic decrease of EPC with increasing doping from strongly underdoped to overdoped samples at high momentum transfer, while closer to the $\Gamma$ point EPC seems to be less affected by screening, consistently with previous calculations \cite{Johnston2010}. 
We have also detected an anomalous dependence of phonon intensity close to the CDW wave vector. However, we have revealed that momentum dependence of the EPC measured at O $K$-edge is different from what measured at Cu $L$-edge \cite{braicovich2020determining}, implying that the RIXS cross section contains additional form factors. The present results show that investigations of the phonon dynamics by O $K$-edge RIXS in presence of delocalized electrons has to be made with precautions in the framework of the models developed so far, and new approaches have to be developed to reach a quantitative reliability of results.

\begin{acknowledgments}
Y.~Y.~P. is grateful for financial support from the Ministry of Science and Technology of China (Grant No. 2019YFA0308401) and the National Natural Science Foundation of China (Grant No. 11974029). P.~A. acknowledges support from the Gordon and Betty Moore Foundation, grant no. GMBF-9452. X.~J.~Z. thank the financial support from the National Natural Science Foundation of China (Grant no. 11888101), the National Key Research and Development Program of China (Grant no. 2016YFA0300300) and the Strategic Priority Research Program (B) of the Chinese Academy of Sciences (Grant no. XDB25000000). S.~J. is supported by the National Science Foundation under Grant No. DMR-1842056. L.~M., M.~M.~S. and G.~G. acknowledge support by the project PRIN2017 “Quantum-2D” ID 2017Z8TS5B of the Ministry for University and Research (MIUR) of Italy. R.A. acknowledges support  by the Swedish Research Council (VR) under the Project 2020-04945. This research used Diamond Light Source beam line I21 under the proposal SP20012. N.~B.~B thanks Diamond Light Source for hosting him at the time the experiment was carried out at I21. This research used the beam line 2-ID of the National Synchrotron Light Source II, a U.S. Department of Energy (DOE) Office of Science User Facility operated for the DOE Office of Science by Brookhaven National Laboratory under Contract No. DE-SC0012704. The work at BNL was supported by the US Department of Energy, oﬃce of Basic Energy Sciences, contract no. DOE-sc0012704.
\end{acknowledgments}

\bibliography{phOK_RIXS}

\end{document}